\documentclass[fleqn,10pt,twocolumn]{wlscirep}

\usepackage[table]{xcolor}
\usepackage[utf8]{inputenc}
\usepackage[T1]{fontenc}

\usepackage{amsmath}
\usepackage{amssymb}
\usepackage{glossaries}
\usepackage{graphicx}
\graphicspath{{draft2_clusters_Figures/}{./}}
\usepackage{here}
\usepackage{times}
\usepackage{units}
\usepackage{float}
\usepackage{booktabs}
\usepackage{bm}

\usepackage{multirow}
\usepackage{makecell}
\usepackage{tabularx}

\usepackage{mathtools}
\usepackage{xspace}
\usepackage{stfloats}
\usepackage{threeparttable}
\usepackage[none]{hyphenat}

\usepackage{caption}
\title{Machine-learning-driven kinetic discovery of carbon interstitial color centers in diamond}

\author[1]{Xiaoya Chang}
\author[2,3]{Nima Ghafari Cherati}
\author[4,5]{Arsalan Hashemi~\thanks{Corresponding author: arsalan.hashemi@uef.fi}}
\author[4,5]{Mikko Karttunen}
\author[2,3,6]{\'Ad\'am Gali~\thanks{Corresponding author: gali.adam@wigner.hun-ren.hu}}
\author[1,7]{Tapio Ala-Nissila~\thanks{Corresponding author: tapio.ala-nissila@aalto.fi}}

\affil[1]{MSP Group, Department of Applied Physics, Aalto University, P.O. Box 15600, FI-00076 Aalto, Espoo, Finland}
\affil[4]{European Laboratory for Learning and Intelligent Systems (ELLIS) Institute Finland, Maarintie 8, 02150 Espoo, Finland}
\affil[5]{Department of Technical Physics, University of Eastern Finland, P.O. Box 1627, FI-70211 Kuopio, Finland}
\affil[2]{HUN-REN Wigner Research Centre for Physics, P.O. Box 49, H-1525 Budapest, Hungary}
\affil[3]{Department of Atomic Physics, Institute of Physics, Budapest University of Technology and Economics, M\H{u}egyetem rakpart 3., H-1111 Budapest, Hungary}
\affil[6]{MTA--WFK Lend\"ulet ``Momentum'' Semiconductor Nanostructures Research Group, P.O. Box 49, H-1525 Budapest, Hungary}
\affil[7]{Interdisciplinary Centre for Mathematical Modelling and Department of Mathematical Sciences, Loughborough University, Loughborough, Leicestershire LE11 3TU, United Kingdom}

\begin{abstract}
Diamond hosts optically active point defects central to quantum technologies, yet carbon self-interstitials introduced during growth and irradiation compete with these centers and form new defects. 
Their configurational landscape remains poorly characterized, because small energy differences decide which minima are stable and how they interconvert. 
Here we construct an interstitial-focused dataset and show that, of three machine-learning interatomic potentials (MLIPs) benchmarked against quantum density functional theory, only MACE reproduces this landscape without misordering competing minima. 
Annealing searches with the validated potential reveal previously unreported di- to octa-interstitial clusters whose formation is not dictated by energetic ordering alone, several of which introduce in-gap states. 
Most notably, a newly identified penta-interstitial, C$_\text{5A}$, exhibits zero-phonon-line-dominated emission with a Debye--Waller factor of 0.70, exceeding those of the interstitial-related 3H (0.45) and IL1 (0.38) centers. 
Thermodynamic screening alone can therefore overlook functional metastable defects, highlighting MLIP-driven kinetic exploration as a route to quantum-defect discovery.

\end{abstract}

\begin{document}

\flushbottom
\maketitle


Diamond is one of the most remarkable materials in Nature, combining exceptional mechanical strength, thermal conductivity, and optical performance~\cite{nie2025microstructure,aharonovich2014diamond,yang2019conductive}.
As demand extends beyond jewelry, the synthesis of laboratory-grown \cite{bundy1955man,balmer2009chemical} diamonds has become a cornerstone of industrial technologies.
While the synthesis routes can produce high quality diamonds, defects remain an inherent consequence of the growth process \cite{nemeth2020complex}.
For example, increasing the CH$_4$ precursor concentration during chemical vapor deposition~\cite{balmer2009chemical} markedly alters diamond morphology: at 5--10\% CH$_4$, films develop a cauliflower-like surface, whereas at 15\% CH$_4$ the microstructure evolves into randomly one-dimensional needle-like grains~\cite{Rakha2010JAP,Sankaran2016_PPP,Sankaran2019_ACS}. These observations underscore how excess carbon can fundamentally reshape defect architectures and crystallinity in diamond.

Importantly, defects do not always undermine the utility of a material. In diamond, point defects can, in fact, be highly beneficial, as they introduce distinctive optoelectronic and magnetic functionalities that are attractive for quantum applications~\cite{pezzagna_quantum_2021,Atature2018NRM,aharonovich2011diamond}.
Among the most prominent are optically active point defects, or color centers, such as the nitrogen-vacancy (NV) and silicon-vacancy (SiV) centers, which can serve as single-photon sources and solid-state qubits~\cite{pershin2025coherence,ashfold2020nitrogen,bezard2024giant}.
Yet, the controlled formation of these centers remains challenging because competing self-interstitials can form and aggregate simultaneously, thereby depleting the constituents required for the target color centers through recombination and trapping processes~\cite{Iakoubovskii2003,Slepetz2014,Dickmann2026MD}.
Unlike carbon vacancies, which are often deliberately introduced, self-interstitials have received far less attention, largely because they are widely assumed to become mobile above 700\,K and recombine upon thermal treatment~\cite{Hunt2000PRB,Kiflawi2007JPCM}.
This highlights the richness and complexity of the self-interstitial defect landscape in diamonds.

Self-interstitials have recently attracted renewed attention, as growing evidence suggests that interstitial-related defects can give rise to quantum sensors, including the TR12 and ST2 centers, which may offer some advantages over the NV center~\cite{Naydenov2009,foglszinger2022tr12,foglszinger2026discoveryST2}.
Although their atomic structures remain unresolved, experimental signatures, including their formation following carbon implantation and electron irradiation, strongly point to interstitial-related defect complexes.
Motivated by these findings, some of us investigated carbon interstitial complexes in computationally tractable supercells.
Our previous study~\cite{cherati2025mono} showed that self-interstitials can host intriguing quantum defects, whereas larger interstitial complexes may be energetically more stable.
However, exploring defect configurational space and obtaining a detailed mechanistic understanding of associated transformations require high-fidelity machine-learned interatomic potentials (MLIPs) tailored to self-interstitial defects.
To the best of our knowledge, no such potential currently exists.
Existing classical potentials and MLIPs~\cite{stuart2000reactive,rowe2020accurate,wang2025density} have not been developed to accurately describe the self-interstitial defect landscape in diamond, where subtle energy differences govern local minima and transition pathways.
Overcoming this bottleneck requires a new MLIP that achieves quantum-mechanical density functional theory (DFT)~\cite{Hohenberg1964PR} level accuracy for energies and forces while retaining transferability across diverse defect configurations.
A related study~\cite{chen2025simulating} has demonstrated the feasibility of this hybrid approach for simulating NV center formation dynamics in the presence of carbon self-interstitials.
Crucially, the configuration of a carbon defect that is observed is frequently determined by formation \emph{kinetics} rather than thermodynamic stability. 
For example, for the carbon-pair (G) center in silicon, kinetic pathways prevent relaxation to the predicted ground state, so the experimentally observed center is kinetically selected and terminal at formation instead of being metastable state relaxing to equilibrium~\cite{deak2023kinetics}.
Capturing such kinetic selection and the corresponding pathways in diamond demands a potential able to follow defect formation and interconversion \emph{dynamically}, not merely to rank static energies.

The accuracy and transferability of an MLIP are determined by its atomic-environment descriptors and the learning architecture.
Most modern frameworks employ many-body descriptors to encode atomic environments.
The Gaussian approximation potential (GAP)~\cite{bartok2010gaussian} employs the smooth overlap of atomic positions (SOAP) descriptor~\cite{Bartok2013-fb} to construct a high-dimensional representation of local environments, enabling structural similarity to be quantified and subsequently learned through kernel regression.
In contrast, the neuroevolution potential (NEP)~\cite{fan2021neuroevolution} adopts a neural-network architecture trained using a separable natural evolution strategy, achieving high computational efficiency owing to its optimized formalism and GPU-accelerated implementation in GPUMD~\cite{xu2025gpumd}.
Unlike the above schemes, message-passing neural networks (MPNNs) represent atomic structures as graphs, with atoms as nodes and interatomic interactions as edges.
By iteratively exchanging and aggregating information between neighboring atoms, they enable each atom to encode progressively longer-range structural information.
As a result, the learned representations capture chemically meaningful features beyond the local cutoff radius.
In particular, within the MPNN framework, the equivariant MACE framework~\cite{batatia2206mace} employs higher-body-order message passing, achieving improved accuracy while remaining scalable and highly parallelizable~\cite{kovacs2023evaluation}.

To accurately model interstitial carbon defects in diamond, we develop a dedicated dataset and benchmark three classes of MLIPs, including GAP, NEP, and MACE, for their ability to capture the energetics and kinetics of stable and metastable defect structures.
Using the optimal model, we perform high-fidelity annealing molecular dynamics (MD) simulations to resolve the point- and cluster-defect landscape of carbon interstitials in diamond, from isolated mono-interstitials to octa-interstitial complexes.
Together with a comprehensive analysis of thermodynamic stability and interconversion kinetics, our results considerably extend the  atomistic picture of carbon interstitial cluster formation and metastability, providing fundamental insights for defect engineering in quantum technologies.

\section*{Results}
\label{sec:results}

\subsection*{MLIP development and validation}
\label{subsec:ml_frameworks}

Since the current carbon MLIP datasets~\cite{rowe2020accurate,wang2025density} are poor in interstitial configurations, we begin by generating a diverse yet compact dataset focused on interstitials.
Our initial dataset was generated by creating a local void through atom removal, followed by randomly inserting additional atoms into the vacant region to generate defect centers.
This strategy deliberately samples high-energy, distorted, and non-equilibrium configurations, capturing regions far from equilibrium and preventing systematic softening of the potential~\cite{deng2025systematic}.
During the active learning stage, distinct configurations are selected using the farthest-point sampling (FPS) method from heating trajectories of defect centers, isolated defects, and previously reported stable defects (Fig.~\ref{fig:workflow}a), ensuring sufficient sampling.
Details of the methodology are provided in Supplementary Note 1 (Dataset construction).
The resulting dataset comprises 2,140 configurations, which were randomly partitioned into training and test sets of 2,033 frames (95\%) and 107 frames (5\%), respectively.
A validation set comprising 16 frames, derived from equilibrium defects reviewed in Ref.~\citenum{cherati2025mono}, is employed to evaluate model performance, while being completely omitted from the training data.
For each frame, DFT-calculated energies, forces, and virials were evaluated for MLIP training.
\begin{table}[htb!]
    \centering
    \caption{\textbf{MLIP models' performance.} Errors for total electronic energy ($E$) and atomic force ($F$) predicted by different MLIP models trained on comparable datasets.
    Energy error in meV/atom; force error in meV/\AA.
    RMSE and MAE represent the root-mean-square error and mean absolute error, respectively, computed for the training (trn), testing (tst), and validation (val) datasets.
    Two published NEP models~\cite{wang2025density, fan2024combining} were additionally evaluated, each trained on distinct datasets.
    In Ref.~\citenum{fan2024combining}, the NEP model was trained using the general carbon dataset developed in Ref.~\citenum{rowe2020accurate}, based on optB88-vdW reference calculations.}
    \label{tab:mlip_error}
    \setlength{\tabcolsep}{2.5pt}
    \renewcommand{\arraystretch}{1.15}
    \resizebox{\columnwidth}{!}{
    \begin{tabular}{ll r rr rr rr}
    \toprule
    Dataset & Model & Error & $E^{\rm{trn}}$ & $F^{\rm{trn}}$ & $E^{\rm{tst}}$ & $F^{\rm{tst}}$ & $E^{\rm{val}}$ & $F^{\rm{val}}$ \\\midrule
    This work & GAP  & \makecell[c]{RMSE\\MAE} &
    \makecell[c]{12.8\\8.4} &
    \makecell[c]{455.9\\276.4} &
    \makecell[c]{13.6\\7.3} &
    \makecell[c]{455.9\\267.7} &
    \makecell[c]{5.2\\4.6} &
    \makecell[c]{175.5\\73.7} \\\midrule
    This work & NEP  & \makecell[c]{RMSE\\MAE} &
    \makecell[c]{4.4\\3.0} &
    \makecell[c]{225.0\\111.4} &
    \makecell[c]{6.6\\3.4} &
    \makecell[c]{222.0\\106.7} &
    \makecell[c]{2.0\\1.8} &
    \makecell[c]{69.1\\27.6} \\\midrule
    This work & MACE & \makecell[c]{RMSE\\MAE} &
    \makecell[c]{2.0\\1.4} &
    \makecell[c]{46.3\\24.1} &
    \makecell[c]{2.1\\1.4} &
    \makecell[c]{109.1\\35.3} &
    \makecell[c]{1.1\\0.9} &
    \makecell[c]{10.2\\4.3} \\\midrule
    Ref.~\citenum{wang2025density} & NEP &
    \makecell[c]{RMSE\\MAE} &
    \makecell[c]{21.8\\17.8} &
    \makecell[c]{373.6\\258.3} &
    \makecell[c]{19.9\\16.2} &
    \makecell[c]{365.2\\248.3} &
    \makecell[c]{3.9\\3.6} &
    \makecell[c]{124.8\\49.6} \\\midrule
    Ref.~\citenum{rowe2020accurate} & NEP\cite{fan2024combining} &
    \makecell[c]{RMSE\\MAE} &
    \makecell[c]{13.5\\10.4} &
    \makecell[c]{263.3\\156.4} &
    \makecell[c]{11.4\\9.1} &
    \makecell[c]{259.6\\151.4} &
    \makecell[c]{2.7\\2.4} &
    \makecell[c]{89.9\\31.9} \\
    \bottomrule
    \bottomrule
    \end{tabular}
    }
\end{table}

To systematically assess how model architecture (Fig.~\ref{fig:workflow}b) influences predictive accuracy, we trained the GAP, NEP, and MACE models on the dataset developed in this work.
Detailed hyperparameter settings are provided in Supplementary Note 2 (\textit{Training of the MLIPs}).
The prediction errors are summarised in Table~\ref{tab:mlip_error} and visualised in Figs.~\ref{fig:workflow}c-d.
The GAP model exhibits the largest error, approximately six times higher than that of MACE, and it struggles to accurately predict interstitial configurations.
This deficiency is not unique to the present study; similar behavior has been reported previously~\cite{rowe2020accurate}.
Although their model performs well for pristine carbon materials, it shows substantial deviations for vacancy and interstitial defects in diamond, with formation energy errors ranging from 25\% to 35\%.
In comparison, the NEP achieves a noticeable improvement in the accuracy of both energy and force predictions compared to the GAP model, consistent with the enhancements reported in Refs.~\citenum{fan2021neuroevolution,fan2022gpumd}.
Interestingly, the state-of-the-art MACE demonstrates exceptional accuracy in energy predictions and a dramatic reduction in force errors.
On the training set, the RMSE for forces is merely 46.3~meV/\AA{} -- only 20\% of the NEP value and 10\% of the GAP error -- highlighting its remarkable performance.
However, this accuracy comes at a higher computational cost, limiting the use of MACE in large-scale and high-throughput simulations.
\begin{figure*}[h!]
    \centering
    \includegraphics[scale=0.73]{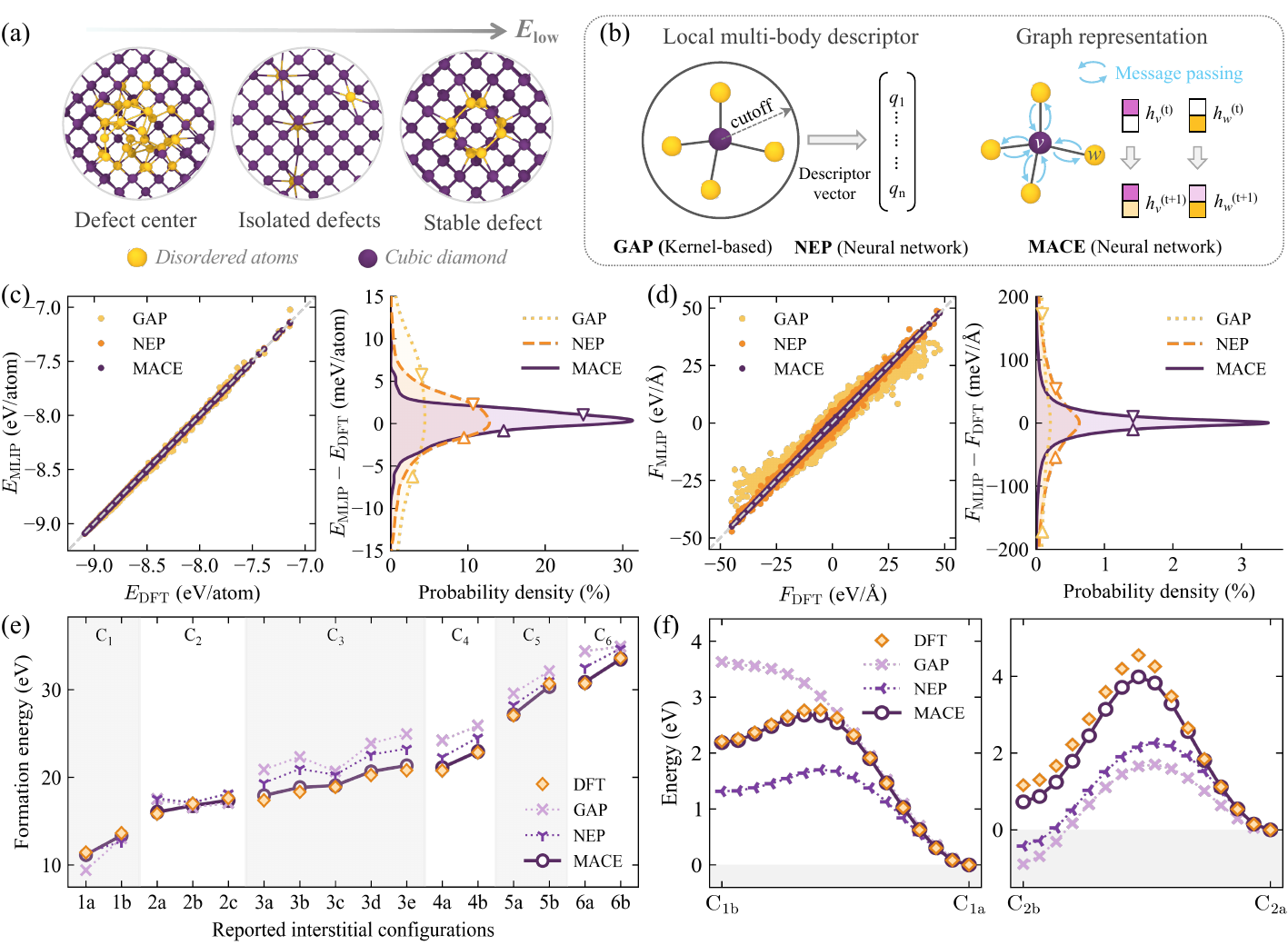}
    \captionsetup{justification=justified}
    \vspace{-1.5em}
    \caption{\textbf{Overview of MLIP development and validation.}
    (a) Schematic illustration of the configurations used for dataset construction.
    (b) Schematic illustration of the descriptors used in the MLIP models.
    (c-d) Comparison of MLIP-predicted energies and force with reference values and the associated error ranges. The lower and upper triangles represent the first and third quartiles of the error distribution, respectively.
    (e) Comparison of formation energies predicted by different MLIP models against the corresponding DFT values for the validation set.
    For each C$_i$ family, structures reported in prior studies are arranged in order of decreasing stability, with the most stable configuration designated as C$_{i,\text{a}}$.
    (f) NEB migration pathways for C$_1$ and C$_2$ calculated by different MLIP models and DFT calculations.
    Here, C$_\text{1a}$, C$_\text{1b}$, C$_\text{2a}$, and C$_\text{2b}$ denote the $\langle 001 \rangle$ split-interstitial, bond-centered, $\pi$-bonded, and Humble configurations, respectively.
    The DFT results in (e-f) were obtained using the same parameter settings for dataset construction in our calculations.}
    \label{fig:workflow}
\end{figure*}

Throughout this study, the defect configuration formed by the incorporation of $i$ interstitial atoms into diamond is denoted as C$_i$.
Each C$_i$ family may comprise multiple equilibrated configurations, which are labeled alphabetically in order of increasing energy; for example, $E(\mathrm{C}_{i\mathrm{a}}) < E(\mathrm{C}_{i\mathrm{b}})$.
This naming follows the previous convention~\cite{cherati2025mono}.
The MLIP frameworks are also assessed for their prediction of the formation energies of charge-neutral defects in the validation set (Fig.~\ref{fig:workflow}e).
The corresponding numerical values are tabulated in Supplementary Table 4.
The MAEs for GAP, NEP, and MACE are 2.37, 1.37, and 0.29 eV, respectively.
Importantly, both GAP and NEP fail to reproduce the correct stability for C$_2$ and C$_3$, indicating limitations primarily in descriptor representation, whereas MACE consistently reproduces both absolute energetics and relative stability.

Nudged elastic band (NEB) calculations were carried out for the mono- and di-interstitial defects (Fig.~\ref{fig:workflow}f).
In this notation, C$_\text{1a}$, C$_\text{1b}$, C$_\text{2a}$, and C$_\text{2b}$ denote the $\langle 001 \rangle$ split-interstitial (two atoms sharing a single lattice site), bond-centered, $\pi$-bonded, and Humble configurations, respectively.
These simulations serve to verify whether the MLIP models can accurately reproduce structural transition barriers and, by extension, defect diffusion.
For the $\mathrm{C}_\text{1b}\rightarrow\mathrm{C}_\text{1a}$ transition, GAP predicts a barrierless pathway, whereas NEP and MACE yield barriers of 0.38 and 0.49 eV, respectively, in close agreement with the DFT value of 0.56 eV.
For the reaction energetics, GAP gives an overestimated value of 3.63 eV, whereas NEP underestimates it to 1.32 eV.
By contrast, MACE predicts 2.19 eV, in excellent agreement with the DFT value of 2.21 eV.
For the more complex $\mathrm{C}_\text{2b}\rightarrow\mathrm{C}_\text{2a}$ transformation, GAP and NEP incorrectly predict $\mathrm{C}_\text{2b}$ to be more stable, in contrast to DFT, whereas MACE closely reproduces the DFT value.
MACE predicts an activation barrier of 3.25~eV, in good agreement with the DFT value of 3.39~eV, whereas GAP and NEP substantially underestimate it to 2.59~eV and 2.68~eV, respectively.
The corresponding reaction energetics are 0.73~eV for MACE and 1.16~eV for DFT, indicating a modest deviation in the MACE prediction.

To demonstrate the need for the diverse dataset constructed in this work, the predictive performance of two NEP models trained on external datasets is also evaluated for energy and force predictions (Table~\ref{tab:mlip_error} and Supplementary Figures 12 and 13) and NEB calculations (Supplementary Figure 14).
Despite sharing the same framework, both models exhibit substantially larger prediction errors on our interstitial dataset and migration pathways, which are attributed to insufficient coverage of defect configurations in their training data.
These highlight the importance of constructing a comprehensive interstitial dataset for reliable modeling of defect systems.

\subsection*{Stable structure screening}
\label{subsec:new_structure}

\begin{figure*}[h]
    \includegraphics[scale=0.56]{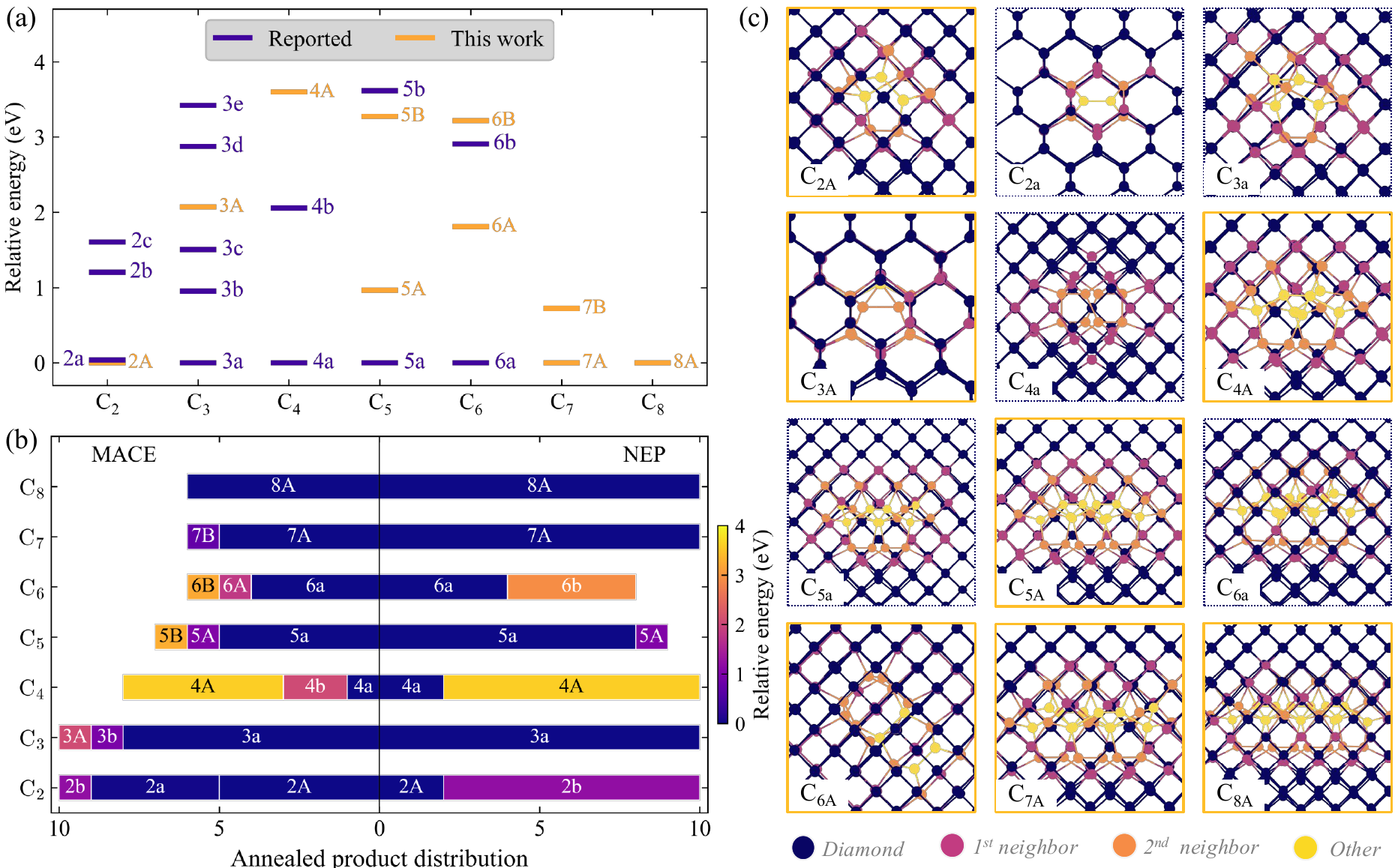}
    \captionsetup{justification=justified}
    \caption{\textbf{Defect structures obtained from MLIP-driven annealing simulations.}
    (a) Relative energies of structures generated from annealing simulations and subsequently optimized by DFT at the PBE level.
    Structures reported by Ref.~\citenum{cherati2025mono} are labeled with lowercase letters (dark purple), while newly discovered configurations are denoted by uppercase letters (orange).
    Within each C$_i$ family, relative energies are defined with respect to the lowest-energy structure.
    (b) Distribution of annealing outcomes from 20 independent simulations, including ten MACE-based and ten NEP-based runs.
    (c) Atomic visualizations of representative structures sampled from the simulations. Atoms are colored by their local structural environments--cubic diamond, first-neighbor, second-neighbor, and other coordination types.
    To ensure reproducibility, all atomic coordinates have been deposited in the \url{Zenodo} repository.
    }
    \label{fig:new_struct}
\end{figure*}
%
Employing validated MLIPs, we explore the defect configurational landscape across a broad range of thermodynamic conditions.
To this end, annealing machine-learned molecular dynamics (MLMD) simulations were performed using the MACE and NEP models to systematically sample point- and cluster-defect structures.
Although NEP is less quantitatively accurate than MACE, its substantially lower computational cost makes it well suited for efficient large-scale simulations, particularly of high-defect-density systems that would otherwise be computationally prohibitive.
All simulation details are provided in the section Methods (\textit{MLMD simulations}).
In this work, we refer to the lowest-energy configuration identified within each defect family as the \textit{stable} structure, while higher-energy local-minimum configurations are referred to as \textit{metastable} structures. 

Initial configurations were generated by randomly placing interstitial carbon atoms throughout the 512-atom diamond lattice, allowing isolated interstitials to migrate and aggregate into more complex defect structures during the annealing simulation.
For each C$_i$ (where $2 \leq i \leq 8$) family, twenty configurations were randomly initiated and subsequently compared using a structure-matching algorithm~\cite{ong2013python} to remove duplicates.
Only final structures lying within 4~eV of the lowest-energy configuration were retained for further DFT optimization, while higher-energy structures were excluded from subsequent analysis.
Figure~\ref{fig:new_struct}a shows the relative energies of the stable structures within each C$_i$ family, referenced to the lowest-energy configuration at PBE level.
Newly discovered configurations in this work are labeled using uppercase letters (e.g., C$_{\text{2A}}$, C$_{\text{4B}}$) to distinguish them from previously reported structures (e.g., C$_{\text{2a}}$, C$_{\text{4b}}$).
For reference, all these configurations were also optimized at the more accurate HSE06 level~\cite{heyd2004efficient}, and their corresponding energies are presented in Supplementary Note 3 (Defect structures).
PBE and HSE06 show consistent trends in the relative energies (Table~S6), except for C$_\text{5A}$, mainly due to the larger singlet--triplet splitting of C$_\text{5a}$ predicted by PBE (Table~S7).
Since charged defects are typically unstable under visible-light illumination due to shallow states and carrier excitation, while neutral defects are more stable across configurations~\cite{cherati2025mono}, we do not explicitly consider charge states in this work.

\begin{figure}[t!]
    \centering
    \includegraphics[width=1\linewidth]{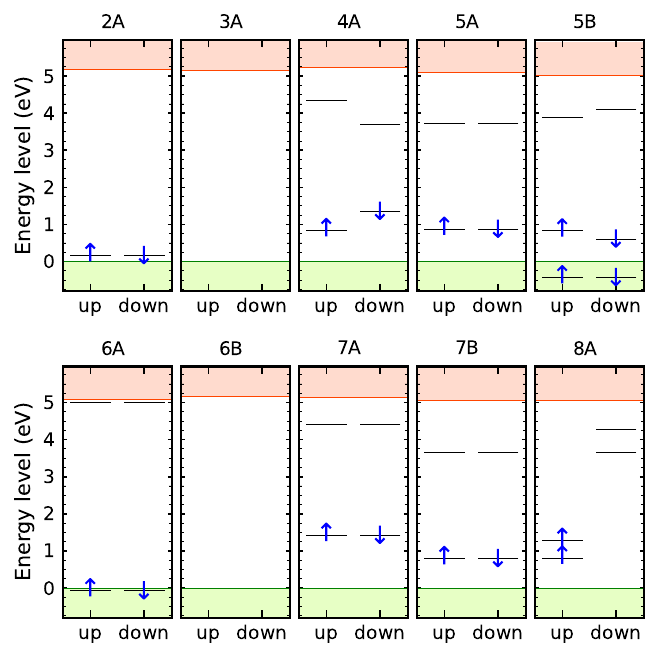}
    \captionsetup{justification=justified}
    \caption{
    \textbf{Ground-state electronic structures of the newly identified defect configurations.}
    The Kohn--Sham (KS) energy levels of the neutral defects were calculated using the HSE06 functional and are shown for the spin-up ($\uparrow$) and spin-down ($\downarrow$) channels. The valence-band (VB) and conduction-band (CB) regions are highlighted in light green and light purple, respectively. The corresponding numerical energy-level values are tabulated in Supplementary Table~S5.
    }
    \label{fig:new_KS}
\end{figure}

Our results reveal a di-interstitial defect, C$_{\text{2A}}$, a distorted ($C_1$ point-group symmetry, with identity only) derivative of the $\pi$-bonded C$_{\text{2a}}$. The energy difference between these two cannot be resolved at the DFT level since 
their ordering is functional-dependent: C$_{\text{2A}}$ lies 37~meV below C$_{\text{2a}}$ at the PBE level on which our dataset is built, whereas HSE06 reverses the order, placing C$_{\text{2a}}$ 14~meV below C$_{\text{2A}}$ (Supplementary Note 3). C$_{\text{2A}}$ is therefore best regarded as a new \emph{metastable} member of the di-interstitial family rather than a lower-energy ground state.
Across the tri- to octa-interstitial families we identify several further new metastable configurations -- C$_{\text{3A}}$, C$_{\text{4A}}$, C$_{\text{5A}}$, C$_{\text{5B}}$, C$_{\text{6A}}$, C$_{\text{6B}}$, C$_{\text{7A}}$, C$_{\text{7B}}$ and C$_{\text{8A}}$ -- whose structural characteristics and formation energies are catalogued in Supplementary Note 3.
Two features matter more than the individual structures.
First, a cage-like tetra-interstitial motif, C$_{\text{4A}}$ ($3.60$~eV at the PBE level and $3.73$~eV at the HSE level above the ground state), recurs as a structural sub-unit of the larger stable clusters (C$_{\text{5a}}$, C$_{\text{6a}}$, C$_{\text{7A}}$, C$_{\text{8A}}$) and, despite its high energy, forms with strikingly high probability (Fig.~\ref{fig:new_struct}b) -- a direct signature of kinetic rather than thermodynamic control.
Second, the new configurations separate cleanly by electronic character: C$_{\text{4A}}$, C$_{\text{5A}}$, C$_{\text{5B}}$, C$_{\text{7A}}$ and C$_{\text{7B}}$ introduce two-level in-gap states of potential interest as color centers (Fig.~\ref{fig:new_KS}; a quantitative assessment will require methods beyond standard DFT~\cite{wasielewski2020exploiting}), whereas C$_{\text{3A}}$ and C$_{\text{6B}}$ are electronically and optically silent.
C$_{\text{6A}}$, an aggregate of C$_{\text{4a}}$ and C$_{\text{2a}}$ units, was previously proposed as a sub-unit of nitrogen--carbon platelets~\cite{Humble_physsci_1982}.

In general, metastable configurations that are local minima on the potential-energy landscape may survive quenching if the barriers separating them from lower-energy configurations are much larger than $k_{\rm B}T$.
Accordingly, several experimentally observed electron paramagnetic resonance (EPR) and optical centers have subsequently been attributed to metastable defect structures generated under non-equilibrium conditions~\cite{twitchen1996electron,twitchen1999optical}.
Notable examples include C$_\text{2b}$, associated with the 3H center~\cite{cherati2025mono}, and C$_\text{2c}$, assigned to the R1 EPR center.
Observed interconversions further suggest that defect evolution is governed primarily by kinetic constraints rather than thermodynamic equilibrium; for instance, direct C$_\text{2c}\rightarrow$ C$_\text{2a}$ transformations can occur without an intermediate C$_\text{2b}$ state, whereas C$_\text{2a}\rightleftharpoons$ C$_\text{2b}$ interconversions become prevalent under strong thermal activation (Supplementary Note 4).
This provides an explanation for the experimental observability of the metastable C$_\text{2b}$ structure (3H center), despite the lower energy of C$_\text{2a}$. 
The relative energies and NEB barriers discussed here are static quantities that neglect entropic contributions, which enter the finite-temperature MD only implicitly. At the annealing temperatures used, a vibrational-entropy difference of one $k_\mathrm{B}$ already shifts relative free energies by $\sim$0.26~eV at 3000~K, exceeding the C$_\text{2A}$-- C$_\text{2a}$ separation; entropic stabilization may therefore contribute alongside kinetics to the observed product distribution.
In practice, some defect or active centers may be reported without atomic-scale resolution; nevertheless, these observations can be correlated with the novel configurations reported here.

Figure~\ref{fig:new_struct}b illustrates the occurrence of the annealed structures generated by the MACE and NEP models.
As a result, the total number of structures is fewer than ten in some cases due to the exclusion of high-energy, highly disordered defective configurations.
For the MACE simulations, the most frequently generated structures are C$_\text{2A}$, C$_\text{3a}$, C$_\text{4A}$, C$_\text{5a}$, C$_\text{6a}$, C$_\text{7A}$, and C$_\text{8A}$.
Among these, the newly identified C$_\text{4A}$ structure is particularly notable: it appears in half of the MACE-based simulations.
By contrast, all other frequently generated structures correspond to the lowest-energy configurations within their respective categories, highlighting the unusual structural prevalence of the higher-energy C$_\text{4A}$ motif.
As visualized in Fig.~\ref{fig:new_struct}c, C$_\text{4A}$ shares a structural motif with the stable C$_\text{5a}$, C$_\text{6a}$, C$_\text{7A}$, and C$_\text{8A}$ structures, characterized by six-membered carbon rings arranged in a zigzag stacking pattern according to the atomic connectivity.
When projected onto the $(011)$ plane, this family of line defects exhibits a characteristic cross-sectional pattern in which six-membered carbon rings are surrounded by six alternating five- and seven-membered rings.
With increasing interstitial content, this chain-like motif progressively elongates while maintaining its geometric coherence, accompanied by localized lattice reconstruction along the chain direction; the elongation of this motif into extended line defects is analyzed elsewhere~\cite{COMPANION_DRAFT1}.

Although trained on similar datasets, different MLIPs may generate distinct potential energy surfaces, leading to variations in diffusion kinetics, energy barriers, and resulting products~\cite{liu2023discrepancies}.
Such method-dependent behavior is evident in the structural distributions obtained from the MACE and NEP simulations shown in Fig.~\ref{fig:new_struct}b, where certain configurations are formed exclusively by one potential.
For instance, the C$_\text{6b}$ structure, proposed as a possible precursor to the TR12 defect~\cite{cherati2025mono}, appears in 40\% of the NEP simulations but is absent in the more accurate MACE results.
This discrepancy likely arises from limited simulations at fixed temperatures (\emph{e.g.}, 4500 K), since the C$_\text{6b}$ structure also appears in MACE simulations under extended sampling conditions (\emph{e.g.}, one out of ten runs at 3000 K).
When comparing the most readily formed defect structures, MACE and NEP show generally consistent behavior, except for the C$_2$ family, where NEP misestimates the relative energetics.
For the remaining cases, the most frequently generated lowest-energy configurations predicted by NEP coincide with those obtained from MACE.
Notably, NEP produces these configurations with a higher probability than MACE, indicating that NEP tends to underestimate energy barriers and thus more readily reaches lowest-energy states, consistent with the C$_1$ and C$_2$ transition behaviors observed in Fig.~\ref{fig:workflow}f.
Overall, our results demonstrate that defect-landscape searches become more effective when multiple computational frameworks are used in combination, with the final candidate set assessed through comparative energetics.

\subsection*{Optical activity of the C$_\text{5A}$ defect}
\label{subsec:5A_str}

To assess whether the characterized defects can exhibit color-center-like features, we select the charge-neutral C$_{\mathrm{5A}}$ defect as a representative case, owing to its well-isolated in-gap states, relatively lower energy, and moderate structural complexity.
Because our model mimics type IIa diamond without donor-acceptor impurities and our previous study~\cite{cherati2025mono} found that neutral configurations are photo-stable and introduce mid-gap states, we limit this analysis to the charge-neutral state.
The localized defect states give rise to multiple, strongly interacting electronic configurations, for which a conventional single-determinant DFT treatment is insufficient.
We therefore employ a multiplet-state description~\cite{Grafenstein2002,Thiering2015,Aleksei2023}, combining group-theoretical analysis with DFT calculations to fully characterize the electronic and optical properties of C$_{\mathrm{5A}}$.

\begin{figure*}[htpb]
    \centering
    \includegraphics[scale=0.96]{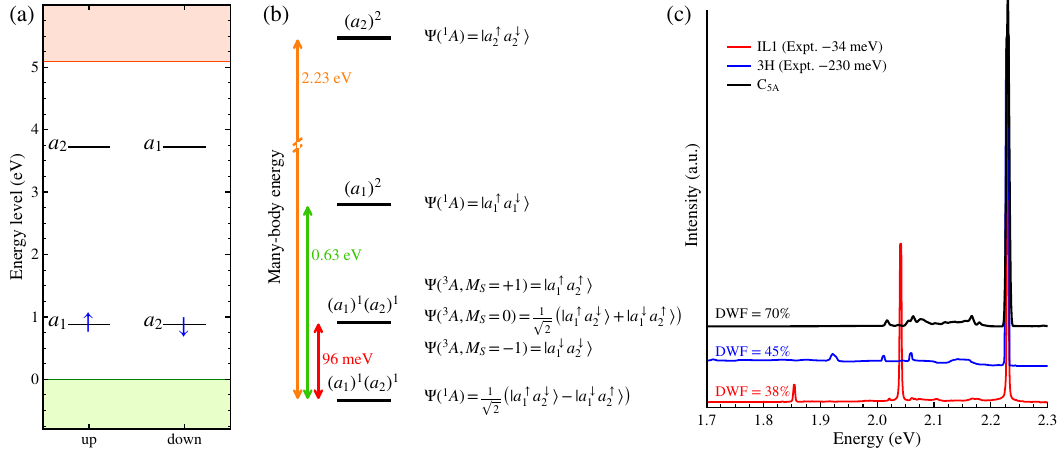}
    \captionsetup{justification=justified}
    \vspace{-0.5em}
    \caption{\textbf{Orbital energies and emission spectra.}
    (a) Kohn-Sham defect levels of neutral C$_{\text{5A}}$ and their orbital notation.
    (b) Many-body states of the C$_{\text{5A}}$ defect and their corresponding wavefunction representations.
    (c) Simulated emission spectrum of C$_{\text{5A}}$ compared with experimental spectra of the IL1 and 3H centers, extracted from Refs.~\citenum{Steeds1999, Sahoo2026}.
    For comparison of the vibronic line shapes, the IL1 and 3H spectra are rigidly shifted by $-34$ and $-230$ meV, respectively, to align their zero-phonon-lines (ZPLs) with that of C$_{\text{5A}}$.
    The spectral alignment is intended only to compare the phonon-side bands and does not imply assignment of C$_{\text{5A}}$ to either experimental center.
    }
    \label{fig:c5A}
\end{figure*}

With $C_1$ symmetry, C$_{\mathrm{5A}}$ is a metastable configuration showing two localized levels within the diamond band gap, denoted $a_1$ and $a_2$ in Fig.~\ref{fig:c5A}a.
For two electrons in two orbitals, the electrons can occupy either the same orbital or two different orbitals.
The corresponding spin-adapted manifold comprises three singlet states and one triplet multiplet.
When the electrons occupy different orbitals, a broken-symmetry calculation can represent the opposite-spin configuration; however, the resulting determinant is not a pure singlet eigenstate of the total-spin operator.
A proper description of the open-shell singlet therefore requires a multiplet-state treatment (Fig.~\ref{fig:c5A}b).
The open-shell singlet ground state is $(a_1)^{1\uparrow}(a_2)^{1\downarrow}$.
The corresponding triplet $(a_1)^{1\uparrow}(a_2)^{1\uparrow}$ lies 96~meV higher.
The closed-shell singlets $(a_1)^2$ and $(a_2)^2$ are about 0.63~eV and 2.23~eV above the ground state, respectively; the latter describes the optically excited singlet (\emph{i.e.}, an electron transition from $(a_1)^{1\uparrow}(a_2)^{1\downarrow}$ to $(a_2)^2$).
Thus, C$_{\mathrm{5A}}$ features a visible-energy singlet transition and a low-lying triplet manifold.

As illustrated in Fig.~\ref{fig:c5A}c, the emission is dominated by the zero-phonon line (ZPL).
The small mass-weighted displacement between ground and excited geometries ($\Delta Q=0.174$ \AA~$\mathrm{amu^{1/2}}$) yields a total Huang--Rhys factor of $S=0.35$ and a Debye--Waller factor (DWF) of $\approx 0.70$, indicative of weak electron--phonon coupling. 
Correspondingly, weak vibronic sidebands peak at $63$, $166$, and $212$~meV from the ZPL.
The calculated transition dipole moment $\mu=4.56$~D yields a radiative decay rate of $\kappa_{\mathrm{rad}}=9.1\times10^{7}$~s$^{-1}$ and a radiative lifetime of $\tau_{\mathrm{rad}}=10.9$~ns.
We attribute the weak electron-phonon coupling to the small redistribution of electronic density during excitation, which produces only minor atomic relaxation between the ground and excited states.

Among self-interstitial-originated color centers reported to date, the 3H center and recently identified IL1 center exhibit experimental DWFs of 0.45 and 0.38, respectively, comparable to that of C$_{\mathrm{5A}}$~\cite{Steeds1999, Sahoo2026}.
The much weaker phonon sideband predicted for C$_{\mathrm{5A}}$ suggests that higher-order interstitial aggregates may also host strongly ZPL-dominated optical transitions.
Although these centers are structurally distinct from C$_{\mathrm{5A}}$, they collectively demonstrate that metastable carbon-interstitial aggregates of varying size can host optically active electronic states.

Several irradiation-induced intrinsic centers reported in the literature remain unidentified, with zero-phonon lines falling in a similar spectral range.
Wang \textit{et al.} observed a ZPL at $550.2$~nm and, based on irradiation and spatial distributions, suggested that several additional centers formed under these conditions are likely interstitial-related~\cite{Wang2017apl}.
Earlier cathodoluminescence studies also found intrinsic radiation-induced centers nearby~\cite{Filipp1992DRM}.
Although these centers are close in energy to the calculated C$_{\mathrm{5A}}$ transition, spectral proximity alone cannot identify them.
A firm assignment would require matching isotope-dependent vibrational signatures, polarization, spin properties, and annealing behavior which is beyond the current study.

We therefore do not assign C$_{\mathrm{5A}}$ to any known optical center.
Nevertheless, its identification through the MLIP-driven configurational search, together with the recovery of several experimentally established defect structures by the same approach, supports the plausibility that C$_{\mathrm{5A}}$ and other shortlisted configurations reported in the SI may also occur in diamond.
More broadly, C$_{\mathrm{5A}}$ illustrates how data-driven exploration of the interstitial configurational landscape can reveal candidate quantum defects without using optical properties as an explicit structural search criterion.

\section*{Discussion}
\label{sec:discussion}

The central challenge in diamond defect physics is not synthesis but atomic-scale identification: carbon scatters weakly, defect clusters are nanoscale and embedded within the three-dimensional bulk, and the configurational landscape is governed by small energy differences that are difficult to resolve experimentally. 
We have shown here that MLIPs provide a quantitative route into this landscape, but that model choice is consequential. 
MACE faithfully reproduces DFT energetics, forces, and migration barriers across our interstitial benchmark, whereas GAP and NEP misorder defect ground states and accumulate errors along aggregation pathways.

Our simulations reveal a series of previously unreported carbon interstitial clusters, extending the known defect landscape of diamond into regions that have remained largely unexplored.
Strikingly, this is true even for the simplest aggregated case: within the extensively studied di-interstitial family, we identify the overlooked C$_\text{2A}$ configuration, which is nearly degenerate in energy with known structures. 
This finding highlights that structurally distinct defect states can remain hidden even in apparently well-characterized parts of the defect landscape. 
More importantly, the fate of these small interstitial clusters is not determined by energetic ordering alone, but by the transition pathways that are kinetically accessible during their formation and evolution. 
This provides a natural explanation for why experimentally observed centers such as 3H (C$_\text{2b}$) and R1 (C$_\text{2c}$) can persist even though C$_\text{2a}$ is the most stable di-interstitial. 
The same principle underlies the kinetically selected G center in silicon~\cite{deak2023kinetics}, reinforcing that searches based only on thermodynamic stability can miss the defect configurations that are actually formed and retained.

The C$_\text{5A}$ defect exemplifies the opportunities enabled by MLIP-driven exploration. 
Identified without prior spectroscopic guidance, it emerges as a particularly promising interstitial color center, with strongly zero-phonon-line-dominated emission and a Debye--Waller factor exceeding those of the established 3H and IL1 centers. 
More broadly, these results demonstrate that high-fidelity MLIPs do not merely accelerate conventional defect searches: they enable a more comprehensive exploration of the defect landscape, revealing metastable configurations, overlooked near-degeneracies, and candidate quantum emitters that may remain inaccessible to searches guided solely by thermodynamic stability or targeted DFT surveys.

\section*{Methods}

\subsection*{DFT calculations}
All DFT calculations used for the construction and labeling of the training dataset were carried out using the Vienna \textit{Ab initio} Simulation Package (VASP)\cite{hafner2008ab} within the projector-augmented wave (PAW) framework.
Exchange-correlation effects were treated using the Perdew--Burke--Ernzerhof (PBE) functional within the generalized gradient approximation (GGA)\cite{Perdew1996-jn}.
The plane-wave basis set was expanded with a kinetic energy cutoff of 550~eV, and electronic self-consistency was achieved with a convergence threshold of $10^{-6}$~eV.
The Brillouin zone was sampled using a Monkhorst--Pack $k$-point mesh with a spacing of 0.2~\AA$^{-1}$.
Spin polarization is taken into account to accurately determine the magnetic ground state and calculate the corresponding structural energy.
Zero-point energy corrections were not included, and all defects were treated in the neutral charge state.
For geometry optimizations, atomic positions were relaxed while keeping the lattice parameters fixed until the total energy change between consecutive ionic steps was below $10^{-6}$~eV and the forces on all atoms were less than 0.02~eV~\AA$^{-1}$.
For selected newly identified defect configurations, additional HSE06 calculations~\cite{heyd2004efficient} were performed to obtain more accurate ground-state electronic structures.


\subsection*{Optical transition calculations}
Emission spectra of C$_\text{5A}$ were calculated within the Huang--Rhys formalism~\cite{Huang1950,alkauskas2014first}, using the adiabatic potential-energy surfaces of the ground and excited electronic states together with the vibrational normal modes coupled to the optical transition. 
Excited-state geometries were obtained using the constrained-occupation $\Delta$SCF approach~\cite{Gali2009}, in which the occupations of the relevant Kohn--Sham orbitals were constrained to preserve the target electronic configuration during ionic relaxation. 
The relaxed ground- and excited-state geometries define the minima of the corresponding potential-energy surfaces and were used to determine the ZPL energy and the structural displacement associated with the optical transition. 
This displacement was projected onto the vibrational normal modes to obtain the mode-resolved Huang--Rhys factors and their total value, which characterize the strength of electron--phonon coupling. 
The total Huang--Rhys factor was further used to determine the Debye--Waller factor, while the mode-resolved contributions were used to construct the phonon sideband and the resulting emission line shape.
Further methodological details are provided in Ref.~\citenum{cherati2025mono}.

\subsection*{MLIP training}
Supplementary Note 1 details the initial data preparation and the iterative expansion of the dataset using active learning.
The active learning procedure was implemented within the NEP framework to iteratively refine the dataset.
The GAP~\cite{bartok2010gaussian} model was trained using the QUIPPY package~\cite{Kermode2020-wu}, incorporating explicit two-body interactions together with \texttt{turbo\_SOAP} descriptors\cite{caro2019optimizing}, which implicitly encode three-body and higher-order effects.
The NEP\cite{xu2025gpumd} model was optimized using full-batch training for 180,000 generations, with an increased energy weight to enhance predictive accuracy.
The MACE\cite{batatia2206mace} model was trained using a two-stage protocol with hidden irreps of ``128x0e + 128x1o'' for 1,000 epochs.
Force loss is prioritized in the first stage, while the energy weight is increased in the second stage to improve convergence and generalization.
Further details on the training hyperparameters for three MLIP models are provided in Supplementary Note 2.

\subsection*{Formation energy and transition barrier}
The interstitial defect formation energy is defined as
\begin{equation}
E_\mathrm{f} = E_{\text{tot}} - \frac{N+n}{N} E_{\text{bulk}},
\end{equation}
where $E_{\text{tot}}$ is the total energy of the supercell containing $n$ interstitial atoms, and $E_{\text{bulk}}$ is the total energy of the pristine supercell with $N$ host atoms.
The NEB path was obtained using ASE\cite{larsen2017atomic} code by interpolating intermediate images between the initial and final minima.
The energies of these images were evaluated using three models and benchmarked against DFT reference calculations.
Notably, neither the intermediate images nor the local minima were directly included in the training dataset.

\subsection*{MLMD simulations}
\label{subsec:md_sim}

To identify stable structures (\textit{Stable structure screening} subsection of the Results), we generated initial defect configurations by inserting two to eight interstitial atoms into a $4\times4\times4$ diamond cubic supercell containing 512 carbon atoms.
The resulting structures were then obtained using a three-stage annealing protocol:
(i) heating the initial configurations containing several isolated defects from 300 K to the target temperature within 1.2 ns --- specifically, 3000 K for C$_2$ to C$_6$ and 4500 K for C$_7$ and C$_8$ --- to provide sufficient kinetic energy to overcome the energy barrier and reach a stable state,
(ii) maintaining the system at the target temperature for 2 ns; and
(iii) cooling down to 1 K over 1.2~ns, followed by a 0.2~ps equilibration.
All simulations were conducted in the $NVT$ ensemble.
Higher applied temperatures increase the internal pressure (roughly 30--50 GPa) simultaneously within the configuration, thereby preventing the system from approaching the liquid or graphitization boundary\cite{marchant2023exploring,luo2022coherent}, as further corroborated by Supplementary Figure~15.
The resulting structures were then subjected to energy minimization using the conjugate gradient (CG) algorithm, yielding locally stable configurations.
Unique structures were identified using a structure-matching algorithm implemented in the pymatgen~\cite{ong2013python} package.
Two structures were considered equivalent when they satisfied the matching criteria with a site-position tolerance of 0.3~\AA and an angular tolerance of $5^\circ$.
Structures lying within 4 eV of the lowest energy configuration were then fully optimized using DFT for further investigation.

All MACE-based MD simulations were performed using the LAMMPS~\cite{thompson2022lammps} software, while the NEP-based simulations were conducted with the GPUMD~\cite{xu2025gpumd} package to achieve higher computational efficiency.
For a 514-atom system, MACE-based MD reaches a performance of 1.85~ns/day on a state-of-the-art NVIDIA H200 GPU (Hopper architecture), while NEP-based simulations with GPUMD achieve 14.87~ns/day on an older Tesla architecture-based NVIDIA V100 GPU.
Atomic visualizations and trajectory generation were performed using the OVITO~\cite{stukowski2010visualization} software, while other MD post-processing was conducted with the MDAPY~\cite{wu2023mdapy} package.

\section*{Acknowledgments}
\label{acknowledgments}

This work was supported by the Academy of Finland through grants 370057 and 373647, and by the European Union and the European Innovation Council through the Horizon Europe project QRC-4-ESP (Grant No. 101129663), and EU Horizon Europe Quest project (No. 10116088).
A.H. and M.K. were supported by the Research Council of Finland (Flagship of Advanced Mathematics for Sensing Imaging and Modeling, grant 358944) and by Foundation PS.
We gratefully acknowledge CSC--IT Center for Science Ltd., Finland, and the Aalto Science--IT project for providing the computational resources and computing time used in this work.
A.G. acknowledges the high performance computational resources provided by KIF$\text{\"{U}}$ (Governmental Agency for IT Development of Hungary) and the European Commission for the projects SPINUS (Grant No.\ 101135699) and QuSPARC (Grant No.\ 101186889).

\section*{Author contributions}
\label{contribution}

\textbf{X. C.} constructed the dataset, developed the MLIP model, performed formal analysis, visualized the results, and wrote the original manuscript draft.
\textbf{N. G. C.} examined the generated structures, performed HSE06 and optical-property calculations, and reviewed and edited the manuscript.
\textbf{A. H.} contributed to conceptualization, methodology design, validation, supervision, and extensively reviewed and edited the manuscript.
\textbf{M. K.} contributed to technical aspects, and reviewed and edited the manuscript.
\textbf{A. G.} critically reviewed the manuscript and provided insightful suggestions.
\textbf{T. A.-N.} supervised the project, acquired funding, provided resources, managed project administration, and reviewed and edited the manuscript.

\section*{Data availability}
\label{data}
All data supporting the conclusions of this study are provided in the Article, Supplementary Information, and associated Source Data files. 
The interstitial training dataset, trained interatomic potential models, and optimized atomic structures will be deposited in the Zenodo repository upon publication.
Source data are provided with this paper.

\section*{Code availability}
The code used for dataset construction, machine-learning interatomic potential training, and data analysis, together with the trained potential models, will be made publicly available on GitHub upon acceptance of the manuscript.

\section*{Competing interests}
The authors declare no competing interests.

\bibliography{main.bib}
\end{document}